\documentstyle[preprint,aps,epsfig,amsbsy,floats]{revtex}
 \tightenlines

\begin{document}
\newcommand{\beq}{\begin{equation}}
\newcommand{\eeq}{\end{equation}}
\newcommand{\beqa}{\begin{eqnarray}}
\newcommand{\eeqa}{\end{eqnarray}}
\newcommand{\sr}{\sqrt}
\newcommand{\fr}{\frac}
\newcommand{\mn}{\mu \nu}
\newcommand{\G}{\Gamma}

\draft \preprint{astro-ph/0409106,~ INJE-TP-04-07}
\title{Inflation parameters from Gauss-Bonnet braneworld}
\author{Hungsoo Kim, Kyong Hee Kim, Hyung Won Lee and  Yun Soo Myung\footnote{E-mail address:
ysmyung@physics.inje.ac.kr}}
\address{
Relativity Research Center and School of Computer Aided Science\\
Inje University, Gimhae 621-749, Korea} \maketitle

\begin{abstract}
We calculate the spectral index and tensor-to-scalar ratio for
patch inflation arisen from the Gauss-Bonnet braneworld scenario.
The patch cosmological models  consist of Gauss-Bonnet (GB),
Randall-Sundrum (RS), and 4D general relativistic (GR) cases. In
order to compare with the observation data, we perform
leading-order calculations for all patch models by choosing
large-field, small-field, and hybrid potentials. We show that the
large-field potentials are sensitive to a given patch model, while
the small-field and hybrid potentials are insensitive to a given
patch model. It is easier to discriminates between quadratic
potential and quartic potential in the GB model rather than RS and
GR models. Irrespective of patch models, it turns out that the
small-field potentials are
 the promising models in view of  the observation.
\end{abstract}

\thispagestyle{empty}
\setcounter{page}{0}
\newpage
\setcounter{page}{1}
\section{introduction}
There has been much interest in the phenomenon of localization of
gravity proposed by Randall and Sundrum (RS)~\cite{RS2}. They
assumed a positive three-brane embedded in the 5D anti de Sitter
(AdS$_5$) spacetime. They obtained a localized gravity on the
brane by fine-tuning a brane tension  to a bulk cosmological
constant. Recently, several authors have studied cosmological
implications of the braneworld scenarios. We wish to mention that
the brane cosmology contains some important deviations from the
Friedmann-Robertson-Walker (FRW) cosmology\cite{BDL1,BDL2}.

On the other hand, it is generally accepted that curvature
perturbations produced during inflation are considered to be
 the origin of  inhomogeneities necessary for
explaining  cosmic microwave background (CMB) anisotropies and
large-scale structures. The WMAP\cite{Wmap1},
SDSS\cite{SDSS,SDSS2}, and other data  put forward more
constraints on cosmological models. These show that  an emerging
standard model of cosmology is the  ${\rm \Lambda}$CDM model.
Further, these results coincide with theoretical predictions of
the slow-roll inflation based on general relativity with a single
inflaton. The latest data \cite{SDSS2} shows a nearly
scale-invariant spectrum with the spectral index
$n_s=0.98^{+0.02}_{-0.02}$,  no evidence of the tensor-to-scalar
ratio  with $R<0.36$, and  no evidence of the running spectral
index with $\alpha_s \simeq 0$.

If the brane inflation occurs, one expects that it provides us
quite different results in the high-energy
region\cite{MWBH,LT,RL,TL,Cal1,KLM1,HL,Cal2}.  Since the
Gauss-Bonnet  term modifies the Friedmann equation at high-energy
significantly,  its application to  the brane inflation has been
studied widely in the
literature\cite{DLMS,Cal3,TSM,KM2,CT,KM3,SST}.

In this work,  the patch cosmological models  induced from the
Gauss-Bonnet braneworld are introduced to study the brane
inflation for large-field, small-field, and hybrid potentials.
We use mainly the leading-order spectral index and
tensor-to-scalar ratio to select which patch model with $q$ is
suitable for explaining the latest observation data.

The organization of this work is as follows. In Section II we
briefly review the patch cosmology arisen from the Gauss-Bonnet
braneworld scenario, and introduce relevant inflation parameters
as observables. We introduce  various potentials to compute their
theoretical values  and compare these with the observation data in
Section III. Finally we discuss our results in Section IV.

\section{patch cosmological models}
We start with an effective  Friedmann equation arisen from the
Gauss-Bonnet brane cosmology by adopting a flat FRW metric as the
background spacetime on the brane\footnote{For reference, here we
add the action for the  Gauss-Bonnet braneworld scenario:
$S=\fr{1}{2\kappa^2_5}\int_{{\rm
bulk}}d^5x\sqrt{-g_5}\Big[R-2\Lambda_5+\alpha\Big(R^2-4R_{\mu\nu}R^{\mu\nu}+
R_{\mu\nu\rho\sigma}R^{\mu\nu\rho\sigma}\Big)\Big]+\int_{{\rm
brane}}d^4x \sqrt{-g}\Big[-\lambda +{\cal L}_{{\rm matter}} \Big]$
with $\Lambda_5=-3\tilde{\mu}^2(2-\beta)$ for an AdS$_5$ bulk and
${\cal L}_{{\rm matter}}$ for inflation. Its exact Friedmann
equation is given by a complicated form:
$2\tilde{\mu}(1+H^2/\tilde{\mu}^2)^{1/2}\Big[3-\beta+3\beta
H^2/\tilde{\mu}^2\Big]=\kappa_5^2(\rho+\lambda)$.}\cite{Cal1,DLMS,TSM}
\begin{equation}
\label{Heq} H^2=\beta^{2}_{q}\rho^{q},
\end{equation}
where $H=\dot{a}/a$, $q$ is a patch parameter labelling different
models and $\beta_{q}^{2}$ is a factor with energy dimension
$[\beta_q]=E^{1-2q}$.  An additional  parameter $\theta=2(1-1/q)$
is introduced for convenience. We call the above defined on the
$q$-dependent energy region as ``patch cosmology". We summarize
three different models and their parameters  in Table I.
$\kappa^{2}_{5}=8\pi/m_{5}^{3}$ is the 5D gravitational coupling
constant and  $\kappa^2_4=8\pi/m_{{\rm Pl}}^{2}$ is the 4D
gravitational coupling constant. $\alpha=1/8g_s$ is the
Gauss-Bonnet coupling with  the string energy scale $g_s$ and
$\lambda$ is the brane tension. Relationships between these are
given by $\kappa^2_4/\kappa^2_5=\tilde{\mu}/(1+\beta)$ and $
\lambda=2\tilde{\mu}(3-\beta)/\kappa_5^2$, where
$\beta=4\alpha\tilde{\mu}^2\ll 1,~\tilde{\mu}=1/\ell$ with AdS$_5$
curvature radius $\ell$. The RS  case of
$\tilde{\mu}=\kappa_4^2/\kappa_5^2$ is recovered when $\beta=0$.
\begin{table}
 \caption{Three relevant models  and their parameters classifying  patch cosmological models.}
 \begin{tabular}{lp{4.5cm}p{3cm}}
 model   & $q~(\theta)$ & $\beta^2_q$ \\ \hline
 GB      & 2/3 ($-1$)   & $(\kappa^2_5/16\alpha)^{2/3}$\\
 RS      & 2 (1)       & $\kappa^2_4/6\lambda$ \\
 GR      & 1 (0)          & $\kappa_4^2/3$
 \end{tabular}
 \end{table}
We have to distinguish between GB ($\beta \ll1$, but $\beta\not=0$
exactly) and RS ($\beta=0$) cases.

Before we proceed, we note that the Gauss-Bonnet braneworld
affects inflation only when the Hubble parameter is larger than
the AdS curvature scale ($H\gg \tilde{\mu}$). As a result, we have
the two patch models of GB case with $q=2/3$ and RS case with
$q=2$ case. For the other case of $H \ll \tilde{\mu}$, one
recovers the 4D general relativistic (GR) case with $q=1$.
Furthermore, we assume that all of AdS curvature scale
$\tilde{\mu}$, Gauss-Bonnet coupling $\alpha$, and brane tension
$\lambda$ are stable, even if the vacuum energy on the brane  is
so large in the high-energy regions that $H\gg \tilde{\mu}$.

Let us introduce  an inflaton $\phi$ whose equation is given by
\begin{equation}
\label{seq} \ddot{\phi}+3H\dot{\phi}=-V^{\prime},
\end{equation}
where dot and prime denote  the derivative with respect to time
$t$ and $\phi$, respectively. The energy density and pressure are
given by $\rho=\dot{\phi}^2/2+V$ and $p=\dot{\phi}^2/2-V$. From
now on, we use the slow-roll formalism for inflation: an
accelerating universe $(\ddot a>0)$ is being derived by an
inflaton slowly rolling down its potential toward a local minimum.
Then Eqs.(\ref{Heq}) and (\ref{seq}) take the following form
approximately:
\begin{equation}
H^2\approx \beta^2_q V^q,~ \dot{\phi}\approx -V'/3H
\end{equation}
which are the  background equations for  patch cosmological
models.  It implies that the cosmological acceleration can be
derived  by a fluid with a vacuum-like equation of state $p
\approx -\rho$. If $p=-\rho$ for  $\dot{\phi}=0$, this corresponds
to a de Sitter inflation with $a(t)=a_0e^{Ht}$. This is a model to
obtain gravitational waves from the braneworld scenario\cite{LMW}.
In order for inflation to terminate and for the universe to
transfer to a radiation-dominated universe, we need  a slow-roll
mechanism. To this end, we introduce Hubble slow-roll parameters
($\epsilon_1,~\delta_n$) and potential slow-roll parameters
($\epsilon_1^q,~\delta_n^q$) as
\begin{equation}\label{srp}
\epsilon_1 \equiv -\frac{\dot H}{H^2}\approx \epsilon_1^q \equiv
\fr{q}{6\beta^2_q}\fr{(V')^2}{V^{1+q}},~~\delta_n \equiv
\frac{1}{H^n\dot{\phi}}\frac{d^{n+1}\phi}{dt^{n+1}}\approx
\delta_n^q
\end{equation}
with\footnote{For another notation, we use
$\delta^q_1=q\epsilon^q_1/2-\eta^q_1$ with $\eta^q_1\equiv
\frac{1}{3\beta^2_q}\frac{V''}{V^q}$.}
\begin{equation}
\delta^q_1=\frac{1}{3\beta^2_q}\Big[\frac{(V')^2}{V^{1+q}}-\frac{V''}{V^q}\Big],~
\delta_2^q = \frac{1}{(3\beta^2_q)^2}\Big[\frac{V'''V'}{V^{2q}}+
\frac{(V'')^{2}}{V^{2q}}-\frac{5q}{2}\frac{V''(V')^{2}}{V^{2q+1}}
+\frac{q(q+2)}{2}\frac{(V')^{4}}{V^{2q+2}}\Big].
\end{equation}
 Here
 the subscript denotes  slow-roll (SR)-order in the slow-roll
expansion. A slow-roll parameter of  $\epsilon^q_1\ge 0$ governs
the equation of states  $p=\omega_q\rho$ with
$\omega_q=-1+2\epsilon^q_1/3q$, which implies that an accelerating
expansion occurs only for
$\epsilon^q_1<1~(\omega_q<-1+2/3q)$\cite{Kinn}.
$\epsilon^q_1=0~(\omega_q=-1)$ corresponds to a de Sitter
inflation. On the other hand, the end of inflation is determined
by $\epsilon^q_1=1~(\omega_q=-1+2/3q)$. Hence the allowed regions
for inflation are different:  $-1 \le \omega<-1/3$ for GR models,
$-1 \le \omega<0$ for GB model, and $-1 \le \omega<-2/3$ for RS
model. If one chooses the inflation potential $V$, then potential
slow-roll parameters ($\epsilon_1^q,~\delta_n^q$) will be
determined explicitly.

We  describe  how  inflation parameters can be calculated using
the slow-roll formalism. Introducing a variable $u^q=a (\delta
\phi^q-\dot{\phi}\psi^q/H)$ where $\delta \phi^q$ is a perturbed
inflaton and $\psi^q$ is a perturbed metric, its Fourier modes
$u^q_k$ in the perturbation theory satisfy the Mukhanov
equation\cite{muk}:
\begin{equation}
\label{eqsn} \frac{d^2u^q_k}{d\tau^2} +
\left(k^2-\frac{1}{z_q}\frac{d^2z_q}{d\tau^2}\right)u^q_k =
0,\end{equation} where $\tau$ is a conformal time defined by
$d\tau=dt/a$, and $z_q=a\dot{\phi}/H$ encodes all information
about a slow-roll inflation for  a  patch model with $q$.
Asymptotic solutions are obtained as
\begin{equation}\label{bc}
u^q_k \longrightarrow \left\{
\begin{array}{l l l}
\frac{1}{\sqrt{2k}}e^{-ik\tau} & \mbox{as} & -k\tau \rightarrow \infty \\
{\cal C}^q_k z_q & \mbox{as} & -k\tau \rightarrow 0.
\end{array} \right.
\end{equation}
 The first solution corresponds to a plane wave on
scale much smaller than the Hubble horizon of $d_H=1/H$
(sub-horizon region), while the second is a growing mode on scale
much larger than the Hubble horizon (super-horizon region). We
consider a relation of  $R^q_{c{\bf k}}=-u^q_{\bf k}/z_q$ together
with $u^q_{\bf k}(\tau)=a_{\bf k}u^q_k(\tau)+a^{\dagger}_{-{\bf
k}}u^{q*}_k(\tau)$. Using a definition of the power spectrum
$P^q_{R_c}(k)\delta^{(3)}({\bf k}-{\bf
l})=\fr{k^3}{2\pi^2}<R^q_{c{\bf k}}(\tau)R^{q \dagger}_{c{\bf
l}}(\tau)>$, one finds the power spectrum for  curvature
perturbations in the super-horizon region
\begin{equation}\label{gps}
P^q_{R_c}(k) = \left(\frac{k^3}{2\pi^2}\right)
\lim_{-k\tau\rightarrow0}\left|\frac{u^q_k}{z_q}\right|^2 =
\frac{k^3}{2\pi^2}|{\cal C}^q_k|^2.
\end{equation}
Our next work is to find unknown coefficients ${\cal C}^q_k$ by
solving the Mukahnov equation (\ref{eqsn}). In general, it is not
easy to find a solution to this equation directly. Fortunately, we
can solve it using either the slow-roll approximation
\cite{SL,STG} or the slow-roll expansion\cite{SG,KLM}. We find the
$q$-power spectrum to the leading-order\cite{KM2}
\begin{equation}
 \label{2ndps}
 P_{R_c}^{q}=\fr{3q\beta^{2-\theta}_q}{(2\pi)^2}\fr{H^{2+\theta}}{2\epsilon_1} \to
 \fr{1}{(2\pi)^2}\fr{H^4}{\dot{\phi}^2},
\end{equation}
where the right hand side should be evaluated at horizon crossing
of $k=aH$.  Using
 $d\ln k \simeq Hdt$,
the $q$-spectral index defined as
\begin{equation}
n_{s}^q(k) = 1 + \frac{d \ln P^{q}_{R_c}}{d \ln k}
\end{equation}
is given   by

\begin{equation}
\label{11} n^{q}_{s}(k) = 1 - 4\epsilon^q_1 - 2\delta^q_1.
\end{equation}
The $q$-running spectral index  is determined, to the
leading-order, by
\begin{equation}
\label{12} \frac{d}{d\ln k} n^{q}_{s}  =
-8\Big(\epsilon^{q}_{1}\Big)^2/q-10\epsilon^q_1\delta^q_1+
2\Big(\delta^{q}_{1}\Big)^2-2\delta^q_2.
\end{equation}
The tensor-to-scalar ratio $R_q$ is defined by \beq
R_q=16\fr{A_{T,q}^2}{A_{S,q}^2},\eeq where the $q$-scalar
amplitude is normalized by \beq A_{S,q}^2=\fr{4}{25}P_{R_c}^{q}.
\eeq The GR($q=1$) tensor amplitude  is given by \beq
A_{T,GR}^2=\fr{1}{50}P_{T,GR} \eeq where
$P_{T,GR}=(2\kappa_4)^2\Big(\fr{H}{2\pi}\Big)^2$ because a tensor
can be expressed in terms of two scalars like $\delta \phi$.
Tensor spectra for GB and RS  are known only for de Sitter brane
with $p=-\rho$\cite{LMW,DLMS}. It implies that tensor calculation
should be limited to the leading-order computation. These are
given by \beq A_{T,q}^2=A_{T,GR}^2F_{\beta}^2(H/\tilde{\mu}), \eeq
where \beq F_{\beta}^{-2}(x)=\sqrt{1+x^2}-\Big(\fr{1-\beta}
{1+\beta}\Big)x^2 \sinh^{-1}\Big(\fr{1}{x}\Big).\eeq In three
different regimes, we approximate $F_{\beta}^2$ as $F_q^2$:
$F_1^2\approx F_{\beta}^2(H/\tilde{\mu} \ll 1)=1$ for GR model;
$F_2^2 \approx F_{\beta=0}^2(H/\tilde{\mu} \gg
1)=3H/(2\tilde{\mu})$ for RS model; $F_{2/3}^2 \approx
F_{\beta}^2(H/\tilde{\mu} \gg 1)=(1+\beta)H/(2\beta\tilde{\mu})$
for GB model. The tensor amplitude up to leading-order is given by
\beq
 A_{T,q}^2=\fr{3q\beta^{2-\theta}_q}{(5\pi)^2}\fr{H^{2+\theta}}{2\zeta_q}
 \eeq
 with $\zeta_1=\zeta_{2/3}=1$ and $\zeta_2=2/3$\cite{Cal3}.
Finally, the tensor-to-scalar ratio is determined by \beq
\label{ttosr}
R_q=16\fr{A_{T,q}^2}{A_{S,q}^2}=16\fr{\epsilon_1}{\zeta_q}\eeq in
the patch cosmological models. Considering  a relation for tensor
spectral index $n_{T}^{q}=-(2+\theta)\epsilon_1$, one finds the
consistency relations  \beq
R_1=-8n_{T}^{1}=16\epsilon_1,~R_2=-8n_{T}^{2}=24\epsilon_1,~R_{2/3}=-16n_{T}^{2/3}=16\epsilon_1.\eeq
The above shows that the RS consistency relation is equal to that
for GR case, but it is different from that for GB case.

 \section{Inflation with  potentials}
 A generic single-field potential can be characterized by two
 energy scales: a height of potential $V_0$ corresponding to the
 vacuum energy density during inflation and a width of the
 potential $\mu$ corresponding to the change of an inflaton
 $\Delta \phi$ during inflation. In general its form  is given by
 $V=V_0f(\phi/\mu)$. Different potentials have different
 $f$-forms. The height $V_0$ is usually fixed by normalization and thus the
 free parameter is just the width $\mu$. We classify potentials
 into three cases: large-field, small-field, hybrid potentials.

 \subsection{Large-field potentials}
 It was shown that the quartic potential of $V=V_0 \phi^4$
 is under  strong observation pressure (ruled out observationally) for GR and RS (GB) models,
  while the quadratic
 potential of $V=V_0\phi^2$ is inside of the 1$\sigma$-bound for GR and GB models  with the range
 of $e$-folding number $50\le
 N\le 60$. This  is obtained from the likelihood analysis based on
 the leading-order calculations to $n_s^q$ and $R_q$ with  patch cosmological models\cite{CT}.
 Here we choose the large-field potentials of $V^{LF}=V_0\phi^p$ with $p=2,4,6$
 for testing these with the patch
cosmology.  In this case   potential slow-roll parameters are
determined by

\begin{eqnarray}\label{LFsl}
\epsilon_1^q &=& {{qp} \over 2}{1 \over [(q-1)p + 2]N + {{qp} \over 2}},\\
\delta_1^q &=& {1 \over 2}{(2 - 2p + qp)  \over [(q-1)p + 2]N + {{qp} \over 2}}. \nonumber \\
\end{eqnarray}
Substituting these into Eqs.(\ref{11}) and (\ref{ttosr}), one
finds two inflation parameters. For a full computation of
inflation parameters, see Ref.\cite{KM3}. The LF-spectral index is
given by
\begin{equation}\label{LFnq}
n_{s}^{LF} =1- \frac{(3q-2)p+2}{[(q-1)p + 2]N + {{qp} \over 2}}
\end{equation}
in the leading-order calculation. The LF tensor-to-scalar ratio
takes the form  \label{LFRq}\beq R^{LF}_q=\frac{8qp}{\zeta_q}
\frac{1}{[(q-1)p + 2]N + {{qp} \over 2}}. \eeq Fortunately, there
is no free parameter for large-field models. The numerical
results\cite{KM3,SDSS} for large-field potentials are shown in
Table II.

\subsection{Small-field potentials}
 In this subsection we choose the small-field potentials of $V^{SM}=V_0[1-(\phi/\mu)^p]$ with
 $p=2,4,6$. Here $\mu$ plays a role of the free parameter. For convenience, we treat $p=2$ and
$p>2$ cases separately. For the $p=2$ case,  potential slow-roll
parameters  are determined by

\begin{eqnarray}
\label{SFp=2}\epsilon_1^q &=& {{q} \over 2}\Big({\phi_f \over \mu}\Big)^2x^q_p e^{-x^q_p N}, \\
\label{SFp=2e}\eta_1^q &=& -{x^q_p \over 2}\Big\{1+q\Big({\phi_f \over \mu}\Big)^2 e^{-x^q_p N}\Big\}  \\
\end{eqnarray}
with a dimensionless parameter $x^q_p \equiv
\frac{2p}{3\beta_q^2\mu^2V_0^{q-1}}$ whose form is given
explicitly  by \beq \label{xval}
 x^{GR}_p=\frac{p}{4\pi}\Big(\frac{m_{{\rm
Pl}}}{\mu}\Big)^2,~x^{RS}_p=\frac{p}{2\pi}\Big(\frac{m_{{\rm
Pl}}}{\mu}\Big)^2\frac{\lambda}{V_0},~
x^{GB}_p=\frac{2p}{3(2\pi)^{2/3}}\Big(\frac{m_{{\rm
Pl}}^{2/3}}{\mu}\Big)^2\Big(\frac{V_0\beta^2}{\tilde{\mu}^2}\Big)^{1/3}.
\eeq
 Here we find useful inequalities: $x^{GR}_p\ll1$ for $\mu \gg
m_{{\rm Pl}}$, $x^{GR}_p\gg 1$ for $\mu \ll m_{{\rm Pl}}$;
$x^{RS}_p\ll1$ for $\lambda/V_0 \to 0$, $x^{RS}_p\gg 1$ for $\mu
\ll m_{{\rm Pl}}$; $x^{GB}_p\ll1$ for $\beta \to 0$, $x^{GB}_p\gg
1$ for $\mu \ll m_{{\rm Pl}}$. This means that the RS model is
obtained from the RS braneworld in high-energy region, while the
GB model is mainly  determined from the Gauss-Bonnet term in the
braneworld. Also the  RS and GB models recover a result of the GR
model in the low-energy limit of $m_{{\rm Pl}}\gg \mu$.
  From a relation for the number of $e$-folding:
$N\simeq-3\beta_q^2\int^{\phi_f}_{\phi}\Big(V^{q}/V'\Big)d\phi$,
one finds a relation, $\phi=\phi_f e^{-x^q_pN/2}$.
  The
SF-spectral index is given by
\begin{equation}\label{SFnq=2}
n_{s}^{SF} =1-6\epsilon^q_1+2\eta^q_1=1-
x^q_p\Big[1+4q\Big({\phi_f \over \mu}\Big)^2 e^{-x^q_p N}\Big]
\end{equation}
in the leading-order calculation. The SF tensor-to-scalar ratio is
\label{SFRq=2}\beq
R^{SF}_q=\frac{16\epsilon^q_p}{\zeta_q}=\frac{8q}{\zeta_q}\Big({\phi_f
\over \mu}\Big)^2x^q_p e^{-x^q_p N}. \eeq We determine $\phi_f$
from a condition of the end of inflation:
$\epsilon^q_1(\phi_f)=1$. For $p=2$ case, one obtains a condition
of $x^q_p<1$ from $n_s^{SF}<1$.  In the case of
$r^q_p=qpx^q_p/4\ll1$, it provides $\phi_f \simeq \mu/(1+q)^{1/p}$
for numerical computation.

In the case of $p>2$, we have different slow-roll parameters
\begin{eqnarray}
\label{SFp>2} \epsilon_1^q &=& \frac{qp x^q_p} {4\Big[\Big({\phi_f
\over \mu}\Big)^{2-p}+
\frac{p-2}{2}x^q_p N\Big]^{\frac{2(p-1)}{p-2}}}, \\
\eta_1^q &=& -\frac{(p-1)x^q_p}{2\Big[\Big({\phi_f \over
\mu}\Big)^{2-p}+
\frac{p-2}{2}x^q_p N\Big]^{\frac{2(p-1)}{p-2}}}-\frac{2(p-1)}{p}\epsilon^q_1. \nonumber \\
\end{eqnarray}
A relation between $\phi$ and $\phi_f$ is given by
$(\phi/\mu)^{2-p}=(\phi_f/\mu)^{2-p}+(p-2)x^q_pN/2$. In general,
the SF-spectral index is given by
\begin{equation}\label{SFnq>2}
n_{s}^{SF} =1- \frac{(p-1)x^q_p}{\Big({\phi_f \over
\mu}\Big)^{2-p}+ \frac{p-2}{2}x^q_p
N}-\frac{5p-2}{8p}\zeta_qR_q^{SF},
\end{equation}
 where the SF tensor-to-scalar ratio is
\label{SFRq>2}\beq R^{SF}_q=\frac{4qpx^q_p}{\zeta_q
\Big[\Big({\phi_f \over \mu}\Big)^{2-p}+ \frac{p-2}{2}x^q_p
N\Big]^{\frac{2(p-1)}{p-2}}}. \eeq Also we get  $\phi_f$ from a
condition of the end of inflation : $\epsilon^q_1(\phi_f)=1$.
However, there is no constraint on $x^q_p$. In the case of
$r^q_p~(x^q_p)\ll1$, we have $\phi_f\simeq \mu/(1+q)^{1/p}$, while
for $r^q_p(x^q_p)\gg1$, we find a connection $\phi_f\simeq \mu
/(r^q_p)^{1/2(p-1)}=\mu/(qpx^q_p/4)^{1/2(p-1)}$. For the case of
$x^q_p\ll1$,  two inflation  parameters are given by \beq
n_{s}^{SF} =1- \frac{(p-1)x^q_p}{(1+q)^{\frac{p-2}{p}}+
\frac{p-2}{2}x^q_p N}-\frac{5p-2}{8p}\zeta_qR_q^{SF}, \eeq and
\label{xSFRq>2}\beq R^{SF}_q=\frac{4qpx^q_p}{\zeta_q
\Big[(1+q)^{\frac{p-2}{p}}+ \frac{p-2}{2}x^q_p
N\Big]^{\frac{2(p-1)}{p-2}}}.\eeq On the other hand, for the case
of $x^q_p\gg 1$, the spectral index is \beq n_{s}^{SF} =1-
\frac{(p-1)x^q_p}{(\frac{qpx^q_p}{4})^{\frac{2-p}{2(1-p)}}+
\frac{p-2}{2}x^q_p N}-\frac{5p-2}{8p}\zeta_qR_q^{SF}, \eeq and the
SF tensor-to-scalar ratio takes the form  \label{xlSFRq>2}\beq
R^{SF}_q=\frac{4qpx^q_p}{\zeta_q
\Big[(\frac{qpx^q_p}{4})^{\frac{2-p}{2(1-p)}}+ \frac{p-2}{2}x^q_p
N\Big]^{\frac{2(p-1)}{p-2}}}. \eeq

In the limit  of $x^q_p\to \infty$, we obtain  the low-energy
limit of GR case from RS and GB models. Also, in the limit of
$x^{GR}_p \to \infty~(m_{{\rm Pl}}\gg\mu)$, one finds a well-known
general relativistic case. These all lead to the same expression
given by \beq n_{s}^{SF} =1- \frac{p-1}{p-2}\frac{2}{N} \eeq which
is independent of the patch parameter $q$. In the limit of $x^q_p
\to \infty$, one finds an asymptotic behavior for the
tensor-to-scalar ratio \label{asSFRq>2}\beq R^{SF}_q \sim
\frac{1}{(x^q_p)^{p/(p-2)}}\to 0. \eeq On the other hand, for
$x^q_p
 \gg1$,
there exist upper limits for $R_p^q$ such that \cite{KKMR}: \beq
R^{SF}_q < \bar{R}^{SF}_q \eeq with
 $\bar{R}^{SF}_q=
R^{SF}_q|_{x^q_p=\bar{x}^q_p}$. Here $\{\bar{x}^q_p\}$=$\{
p/4\pi,~(p/2\pi)(\lambda/V_0),~[2p/3(2\pi)^{2/3}](V_0
\beta^2/\tilde{\mu}^2)^{1/3} \}$ is the $x^q_p$-value for $m_{{\rm
Pl}}=\mu$ in Eq.(\ref{xval}). The numerical results for $x^q_p
\gg1$ \cite{SDSS} and those from graphical analysis for $x^q_p
\ll1$\cite{TL}  are summarized at the last column in Table II.

\begin{table}
\caption{The spectral index ($n_s$) and tensor-to-scalar ($R$).
Here we choose $N=55$ to find theoretical values for the
large-field potentials (LF) and bounds for small-field potentials
(SF). For SF case, each patch model in high-energy region is
recovered when $x^q_p\ll1$, while their low-energy limits are
recovered when $x^q_p\gg 1$. The  patch cosmological model is
allowed only for $x^q_p\ll1$ because for $x^q_p\gg 1$, it
degenerates GR case.}
 \begin{tabular}{|c|c|c|c|c|}
   Patch & $p$ & LF& SF($x^q_p\ll 1$) & SF($x^q_p\gg 1$)\\\hline
         &     & $n_s=0.97 $ & $n_s\le 1$          & N/A  \\
         &2& $R = 0.14 $ & $R \le 0.04$ & N/A  \\ \cline{2-5}
     GB  & & $n_s=0.95 $ & $n_s\le 1$          & $0.95\le n_s\le 1$ \\
$(q=2/3)$ &4& $R = 0.56 $ & $R\le 1.9\times 10^{-3}$& $R\le \bar{R}_{2/3}^4$ \\
\cline{2-5}
               & &    N/A      & $n_s\le 1$   & $0.96  \le n_s\le 1$ \\
               &6&    N/A      & $R\le 6.0\times 10^{-4}$& $R\le \bar{R}_{2/3}^6$ \\ \hline

               & & $n_s=0.96 $ & $n_s\le 1$   & N/A  \\
               &2& $R = 0.14 $ & $R\le 0.05$& N/A  \\ \cline{2-5}
GR     & & $n_s=0.95$ & $n_s\le 1$& $0.95\le n_s\le 1$  \\
$(q=1)$     &4& $R = 0.29$ & $R\le 1.3\times 10^{-3}$ & $R\le 9.5\times 10^{-4}$  \\
\cline{2-5}
               & & $n_s=0.93$ & $n_s\le 1$ & $0.96\le n_s\le 1$  \\
               &6& $R = 0.43$ & $R\le 3.0\times 10^{-4}$ & $R\le 5.7\times 10^{-4}$  \\ \hline

               & & $n_s=0.96$ & $n_s\le 1$& N/A  \\
               &2& $R = 0.22$ & $R\le 0.16$ & N/A  \\ \cline{2-5}
RS        & & $n_s=0.95$ & $n_s\le 1$ & $0.95\le n_s\le 1$  \\
$(q=2)$     &4& $R = 0.29$ & $R\le 8.0\times 10^{-4}$ & $R\le \bar{R}_2^4$  \\
\cline{2-5}
               & & $n_s=0.94$ & $n_s\le 1$ & $0.96\le n_s\le 1$  \\
               &6& $R = 0.32$ & $R\le 1.0\times 10^{-4}$ & $R\le \bar{R}_2^6$\\
\end{tabular}
\end{table}

\subsection{Hybrid potentials}
Finally we choose the hybrid-field potentials (HY) like
$V^{HY}=V_0[1+(\phi/\mu)^p]$ with $p=2,4,6$.  In this case it
requires an auxiliary field to end inflation. Here we separate
$p=2$ and $p>2$ cases.

For the $p=2$ case, the potential slow-roll parameters  are
determined by
\begin{eqnarray}
\label{HYp=2}\epsilon_1^q &=& {{q} \over 2}\Big({\phi_f \over \mu}\Big)^2x^q_p e^{x^q_p N}, \\
\label{HYp=2e}\eta_1^q &=& {x^q_p \over 2}\Big\{1-q\Big({\phi_f \over \mu}\Big)^2 e^{-x^q_p N}\Big\}  \\
\end{eqnarray}
with a dimensionless parameter $x^q_p=
2p/3\beta_q^2\mu^2V_0^{q-1}$ defined in Eq.(\ref{xval}).
 From
$N \simeq -3\beta_q^2\int^{\phi_f}_{\phi} (V^{q}/V')d\phi$, one
finds  a relation, $\phi=\phi_f e^{x^q_pN/2}$ for $p=2$. Here
$\mu$ and $\phi_f$ are regarded  as  free parameters. The
HY-spectral index is then given by
\begin{equation}\label{HYnq=2}
n_{s}^{HY} =1-6\epsilon^q_1+2\eta^q_1=1+
x^q_p\Big[1-4q\Big({\phi_f \over \mu}\Big)^2 e^{x^q_p N}\Big]
\end{equation}
in the leading-order calculation. The HY tensor-to-scalar ratio is
found to be  \beq \label{HYRq=2}
R^{HY}_q=\frac{16\epsilon^q_p}{\zeta_q}=\frac{8qx^q_p
}{\zeta_q}\Big({\phi_f \over \mu}\Big)^2e^{x^q_p N}. \eeq In order
that Eqs.(\ref{HYnq=2}) and (\ref{HYRq=2}) be meaningful, we
require a condition of $x^q_p<1$. On the other hand, there is no
way to determine $\phi_f$ from $\epsilon^q_1(\phi_f)=1$ for HY
case because $\phi_f$ is determined by  other mechanism. Hence
$\phi_f$ plays a role of  the  free parameter.  Fortunately,
$|(\phi/\mu)|^2<1$ is required because if $|(\phi/\mu)|^2>1$ in
$V^{HY}$, it is not much different from the large-field
potentials. From Eq.(\ref{HYnq=2}), one finds a  restrictive
constraint  $|(\phi/\mu)|^2<\frac{1}{4q}$ which  comes from  the
condition of $n_s^{HY}>1$.

In order to see a feature of the hybrid models, we need  numerical
results. In the case of $x^{GR}_{p=2}=0.04~(\mu=2m_{{\rm Pl}}),
~\ln(\phi/\phi_f)= 1$ and $N=50$, we have  a blue spectral index
$n_s^{HY}\simeq 1+x^q_p=1.04$ but a small tensor-to-scalar ratio
$R^{HY}_{GR}=0.32(\phi/\mu)^2<0.08$.

 In the case of $p>2$,
we have different slow-roll parameters
\begin{eqnarray}
\label{HYp>2} \epsilon_1^q &=& \frac{qp x^q_p} {4\Big[\Big({\phi_f
\over \mu}\Big)^{2-p}-
\frac{p-2}{2}x^q_p N\Big]^{\frac{2(p-1)}{p-2}}}, \\
\eta_1^q &=& \frac{(p-1)x^q_p}{2\Big[\Big({\phi_f \over
\mu}\Big)^{2-p}-
\frac{p-2}{2}x^q_p N\Big]^{\frac{2(p-1)}{p-2}}}-\frac{2(p-1)}{p}\epsilon^q_1. \nonumber \\
\end{eqnarray}
In general, the HY-spectral index is given by
\begin{equation}\label{gHYnq>2}
n_{s}^{HY} =1+\frac{(p-1)x^q_p}{\Big({\phi_f \over
\mu}\Big)^{2-p}- \frac{p-2}{2}x^q_p
N}-\frac{5p-2}{8p}\zeta_qR_q^{HY},
\end{equation}
 where the HY tensor-to-scalar ratio is
\label{gHYRq>2}\beq R^{HY}_q=\frac{4qpx^q_p}{\zeta_q
\Big[\Big({\phi_f \over \mu}\Big)^{2-p}- \frac{p-2}{2}x^q_p
N\Big]^{\frac{2(p-1)}{p-2}}}. \eeq A HY-relation for $ p>2$ is
given by $(\phi/\mu)^{2-p}=(\phi_f/\mu)^{2-p}-(p-2)x^q_pN/2>0$,
which implies that  an inequality of $N_{{\rm max}}>N$  exists
with the definition of $N_{\rm max}$ as $(\phi_f/\mu)^{2-p} \equiv
(p-2)x^q_pN_{{\rm max}}/2$.

Consequently, the HY-spectral index is given by
\begin{equation}\label{HYnq>2}
n_{s}^{HY} =1+ \frac{p-1}{p-2}\frac{2}{N_{{\rm
max}}-N}-\frac{5p-2}{8p}\zeta_qR_q^{HY},
\end{equation}
 where the HY tensor-to-scalar ratio is
\label{HYRq>2}\beq R^{HY}_q=\frac{4qp}{\zeta_q
(x^q_p)^{\frac{p}{p-2}}\Big[\frac{(p-2)N_{{\rm
max}}}{2}\Big]^{\frac{2(p-1)}{p-2}}}\frac{1}{\Big[1-\frac{N}{N_{{\rm
max}}}\Big]^{\frac{2(p-1)}{p-2}}}. \eeq

\section{discussions}

We introduce  various potentials which are classified into
large-field, small-field, and hybrid types for  the ordinary
inflation in the GR case. Using the patch cosmological models
together with various potentials, we compute the two cosmological
observables, spectral index and tensor-to-scalar ratio.

In large-field models without free parameter, the spectral index
$n_s$ and tensor-to-scalar ratio $R$ depend on the $e$-folding
number $N$ only. Actually this simplicity provides strong
constraints on large-field models. Further, combining the
Gauss-Bonnet braneworld with large-field potentials provides more
tighten constraints than the 4D general relativistic case. The GB
 case is regarded as the promising model for testing the
large-field potentials because it accepts the quadratic potential.
On the other hand, it rejects the quartic potential because
theoretical points are far outside the $2\sigma$-bound\cite{CT}.
Actually, the GB cosmological model improves the theoretical
values predicted by the GR model, whereas the RS model provides
indistinctive values more than the GR case.  As is shown in Table
II, the GB model splits large-field potentials into three distinct
regions clearly: for $N=55$, $n_s^{GB}=0.97 \to 0.95~(p=2\to
p=4),~R_{GB}=0.14 \to 0.56$, and a power-law inflation with $p=6$:
$n_s^{PI}=1-[(2r-1)/2r^4]^{1/3},~R_{PI}=16/r$\cite{KM2}.
Contrastively, we have $n_s^{GR}=0.96 \to 0.95 \to 0.93~(p=2\to
p=4\to p=6),~R_{GR}=0.14 \to 0.29 \to 0.43$, whereas
$n_s^{RS}=0.96 \to 0.95 \to 0.94,~R_{RS}=0.22 \to 0.29 \to 0.32$.
Theoretical points predicted by the RS model  lie very close to
the border between the regions allowed and disallowed by
observation. Consequently, the large-field models depend on
critically which model is used for calculation.

In small-field potentials, one has a free parameter $\mu~(x^q_p)$
related to the potential shape. It is thus difficult to constrain
inflation parameters, in compared to the large-field potentials.
However, combining the graphical analysis with the data\cite{TL},
we find useful bounds.  For $x^q_p \ll 1$, there is no constraint
on the lower-bound for the spectral index, but all of its upper
bounds are given by 1. This means that patch cosmological model is
not useful for testing  small-field potentials. For $x^q_p \gg 1$
and $p>2$, one finds lower-bounds for the spectral indices.
Furthermore  there is no constraint on the spectral indices for
$p=2$ case. This implies that although $n_s$ is independent of the
patch parameter $q$, the small-field potentials  are in good
agreement with the observational data\cite{SDSS,SDSS2}. Combining
the Gauss-Bonnet braneworld with small-field potentials, there
exist unobserved differences in the  upper-bound of the
tensor-to-scalar ratio $R$.

Concerning the hybrid models, we find a blue spectral index with
$n_s^{SF}>1$. However, there is no actual difference between patch
cosmological model because one more free parameter is necessary to
determine  the end of inflation ($\phi_f$), in addition to $\mu~(
x^q_p)$. It implies that the scheme of  inflation ($q$) is less
important than mechanism of the hybrid inflation ($\mu,\phi_f$).
As a result, we do not find any new result when combining the
Gauss-Bonnet braneworld with the hybrid potentials.

In conclusion, the GB model is still a  promising one to
discriminate between the quadratic and quartic potentials in the
large-field type  by making use of  the observation data. Although
the small-field potential are insenstive to patch cosmological
models, these are considered as the  promising potentials in view
of the observational data.  Finally we do not find  any new result
when combining the Gauss-Bonnet braneworld with  the hybrid
potentials. This implies that it is not  easy for hybrid type to
compare with the  data\cite{SDSS,SDSS2}.

\subsection*{Acknowledgements}
 Y.S. was supported in part by KOSEF, Project No.
R02-2002-000-00028-0. H.W. was in part supported by KOSEF,
Astrophysical Research Center for the Structure and Evolution of
the Cosmos.

\end{document}